\documentclass[useAMS,usenatbib]{mn2e}
\usepackage{graphicx}

\title[Hybrid Mechanism Forming 2:1 Librating-Circulating Configuration]
{A Hybrid Mechanism Forming a 2:1 Librating-Circulating Resonant
Configuration in the Planetary System}

\author[Niu Zhang, Jianghui Ji and Zhao Sun]{Niu Zhang$^{1,2}$, Jianghui
Ji$^{1}$\thanks{E-mail:jijh@pmo.ac.cn (JHJ)} and Zhao Sun$^{1,2}$\\
$^{1}$Purple Mountain Observatory, Chinese Academy of Sciences,
Nanjing 210008, China\\
$^{2}$Graduate School of Chinese Academy of Sciences, Beijing
100049, China}

\begin{document}



\maketitle


\begin{abstract}
A diversity of resonance configurations may be formed under
different migration of two giant planets. And the researchers show
that the HD 128311 and HD 73526 planetary systems are involved in a
2:1 mean motion resonance but not in apsidal corotation, because one
of the resonance argument circulates over the dynamical evolution.
In this paper, we investigate potential mechanisms to form the 2:1
librating-circulating resonance configuration.

In the late stage of planetary formation, scattering or colliding
among planetesimals and planetary embryos can frequently occur.
Hence, in our model, we consider a planetary configuration of two
giants together with few terrestrial planets. We find that both
colliding or scattering events at very early stage of dynamical
evolution can influence the configurations trapped into resonance. A
planet-planet scattering of a moderate terrestrial planet, or
multiple scattering of smaller planets in a crowded planetary system
can change the resonant configuration. In addition, collision or
merging can alter the masses and location of the giant planets,
which also play an important role in shaping the resonant
configuration during the dynamical evolution. In this sense, the
librating-circulating resonance configuration is more likely to form
by a hybrid mechanism of scattering and collision.

\end{abstract}


\section{Introduction}
To date, more than 40 multiple planetary system have been detected
beyond our solar system. The commensurability of orbital periods is
very ubiquitous in the extrasolar planetary systems. At present,
four resonant pair of planets (GJ 876, HD 82943, HD 128311, HD
73526) are reported to be trapped in 2:1 mean motion resonance
\citep{mar01, may04, vog05, tin06}. In recent years, numerous
researchers have extensively investigated the dynamics and origin of
the 2:1 resonance in the planetary systems \citep{Goz01, lee02,
Haj02, Haj06, Haj03, Ji03a, Ji03b, lee04, Kle04, Bea03, Bea06,
Gay08, lee09, voy09}. The two resonance variables for 2:1 resonance,
$\theta_1=\lambda_1-2 \lambda_2+\varpi_1$, $\theta_2=\lambda_1-2
\lambda_2+\varpi_2$ (where $\lambda$, $\varpi$ are the mean
longitude and the longitude of periapse respectively, the subscripts
1, 2 denote the inner and outer planets), are categorized to librate
about: (1)0$^\circ$ in symmetric configuration, (2) both 0$^\circ$
and 180$^\circ$ respectively, in the so-called antisymmetric
configuration, and (3) other degrees different from 0$^\circ$ or
180$^\circ$ in asymmetric configuration \citep{Haj02, Ji03a, Ji03b,
Ji03c,  Bea03, Bea06, lee04}. The GJ 876 was revealed to be the
first 2:1 resonant system \citep{mar01} near an M dwarf star. The GJ
876 system is in apsidal corotation where the mean motion resonance
variables, $\theta_1$ and $\theta_2$ librate about $0^\circ$ with
quite slight amplitudes. On the origin of mean motion resonances in
the system, a formation scenario is that they were assembled by
migration of planets. In the formation of giant planets, if two
planets are massive enough to open gaps and not far away from each
other in the disk, the material between the region of them can be
rapidly cleared off. And then, the dissipation of the stuff outside
the outer planet and inside the inner planet may still force two
planets approach each other. Any process that makes two bodies
approach each other, which are originally separated appropriately,
could result in mean motion orbital resonance \citep{lee02, Kle05,
mas06}.

Lee and collaborators explored the origin and diversity of the 2:1
mean motion resonance \citep{lee02, lee04, Kle05, lee06}. They set
the two planets on coplanar, circular orbits with, $a_2/a_1\sim2/1$.
The outer planet can migrate toward the center star at a reasonable
velocity (\citealt{war97}; see also \citealt{lee02} and references
therein),
\begin{eqnarray}
\left|\frac{\dot{a}}{a}\right|\approx\frac{3\nu}{2a^2}
=\frac{3}{2}\alpha\left(\frac{H}{a}\right)^2\Omega
=9.4\times10^{-5}\left(\frac{\alpha}{4\times10^{-3}}\right)\nonumber\\
\times\left(\frac{H/a}{0.05}\right)^2
\left(\frac{M_\star}{1M_{\odot}}\right)^{1/2}
\left(\frac{a}{1\rmn{AU}}\right)^{-3/2}~\rmn{yr}^{-1}
\end{eqnarray}
where $\nu=\alpha H^2\Omega$, is the kinematic viscosity expressed
using the Shakura-Sunyaev $\alpha$-prescription, while $H$,
$\Omega=2\pi/P$ are the height scale of the protoplanetary disk and
the orbital angular velocity at $a$ (where $P\approx 2\pi
a^{3/2}(GM_\star)^{-1/2}$), and $M_\star$ is the mass of the center
star. In  $(1)$, the values of $\alpha = 4\times 10^{-3}$ and
$H/a=0.05$ are typical in classical models of protoplanetary disks.
If we adopt $M_\star=1M_\odot$, the migration velocity of the planet
with $a=1\rmn{AU}$, is $9.4\times 10^{-5}~\rmn{yr}^{-1}$ by $(1)$.
They find that such migration of the planets can result in resonance
capture with the eccentricities of the planets growing quickly. If
the eccentricity damping induced by planet-disk interaction is
considered, the system may remain stable over secular timescale
after the resonance capture has happened. In such circumstances, the
eccentricities reach nearly constant values. The investigations
imply that the 2:1 symmetric resonant configurations may be easily
established under the forced migration due to planet-disk
interaction.

The numerical explorations and theoretical analysis show that the
2:1 resonance planetary configurations could be quite diverse due to
different migration in a slightly eccentric disk \citep{lee04,
Kle05}. For example, it is shown that the orbital solutions of the
resonant pairs derived from Keplerian fit, could result in unstable
behaviors \citep{vog05, tin06} over the timescale of several
thousand years in the HD 128311 and HD 73526 systems. The best-fit
dynamical orbital solutions are given in Table 1, which each system
can remain stable over $10^4$ yr. Indeed, both of the systems are
stabilized by the 2:1 resonance, however, with a
librating-circulating resonance configuration. Planetary
configurations such as listed in Table 1 are not observed to lead to
the convergent migration scenario mentioned by Lee et al. (2002,
2004, 2006). However, \citet{tin06} mentioned several possibilities
of the origin of librating-circulating resonance configuration. The
resonance configurations may be formed either through rapid
migration or migration with initial planetary eccentricities or via
a dynamical scattering event. \citet{san06} provide a mixed
evolutionary scenario for such resonance configuration by combining
an adiabatic migration progress and a sudden gravitational
perturbation. In their study, firstly, they show the planets could
be captured into mean motion resonances via inward migration. And
then, they describe a scenario of a sudden stop of inward migration
after the resonance capture happened, which is supported by
observations of young protoplanetary disk \citep{cal05, dal05,
mas06}. Finally, a small body ($\sim 10M_\oplus$) is scattered
during the follow-up dynamical evolution of the system. Therefore,
the resonance configuration of the system is turned into a
librating-circulating stage, and the eccentricities of the planet
oscillate with large amplitudes. The results of the numerical
studies are consistent with the dynamical behaviors expressed using
solutions listed in Table 1 of HD 128311 and HD 73526 planetary
systems \citep{san06, san07}.

\begin{table}
\centering
\begin{minipage}{100mm}
\caption{Dynamical orbital fits for two systems.}
\begin{tabular}{@{}ccccccc}
\hline & \multicolumn{2}{c}{HD 128311\footnote{The orbital
parameters reference to  \citet{san06}}} & \multicolumn{2}{c}{HD
73526\footnote{\citet{tin06}}} & \multicolumn{2}{c}{HD
73526\footnote{\citet{san07}}}\\
Parameter & b & c & b & c & b & c \\
\hline
$\rmn{Mass}(M_J)$         & 1.56  & 3.08  & 2.9  & 2.5  & 2.415 & 2.55   \\
$a(\rmn{AU})$......       & 1.109 & 1.735 & 0.66 & 1.05 & 0.659 & 1.045  \\
$e$................       & 0.38  & 0.21  & 0.19 & 0.14 & 0.26  & 0.1107 \\
$\varpi(deg)$......       & 80.1  & 21.6  & 203  & 13   & 202.9 & 253.7  \\
$M(deg)$......            & 257.6 & 166.0 & 86   & 82   & 70.7  & 170.7  \\

\hline
\end{tabular}
\end{minipage}
\end{table}

The terrestrial planets are rocky planets, with masses ranging in
$1-10M_\oplus$. The observations show that there are several systems
containing several super Earths in close-in orbits, e.g., Gl 581 and
HD 40307 (see http://exoplanet.eu/), and recently the nearby
solar-like star 61 Virginis was reported to harbor two Neptune-like
planets at about 0.22 AU and 0.48 AU \footnote{The terrestrial
planets with masses as small as 2 Earth mass and a period of several
months orbiting a K dwarf may be detected in the Habitable zones, by
observing 60 groups of 10 nights with HARPS on the ESO 3.6 m
telescope (M. Mayor 2009, private communication). Ground-based
microlensing is sensitive to Earth-mass planets orbiting at $~1$ AU
(D. Bennett 2009, private communication) around late-M stars, while
MPF is sensitive to planets down to 0.1 Earth mass in the Habitable
zone (near 1 AU) of solar-type, G and K stars. The future E-VLTs and
space missions (e.g., GAIA and SIM) will be also hopeful to discover
multiple Earth-like planets residing in the terrestrial region in a
planetary system.}, and a super Earth \citep{vog09}. This indicates
that terrestrial or Neptune-like planets are very common to survive
in the late stage of planet formation. On the other hand, in the
numerical studies on the terrestrial planet formation, $2-4$
terrestrial planets are formed with moderate eccentricities
\citep{cha01, ray04, ray05, ray06, Zha09}.  It is worthy to pay
attention to direct or indirect influence upon the resonance
configuration in the very beginning of evolution. At that time,
giant planets have been already formed and stopped migrating, while
the terrestrial planets may have been created but unstable. If the
periods of the giants are commensurable, the terrestrial planets
with rather moderate eccentricities could take effect on the
dynamical evolution of a system. For example, the earlier works
\citep{san06, san07} show that librating-circulating configurations
of the 2:1 mean motion resonance can be formed by combining
migration processes and sudden perturbations. In this work, we
investigate potential mechanisms of making a librating-circulating
resonance configuration. And a novel hybrid mechanism is proposed to
explain this issue.  In the model, we consider the planetary
configuration of the two giants together with few terrestrial
planets. We show that a scattering of a moderate mass planet or
multiple continuous scattering of the terrestrial bodies can form
librating-circulating configuration. Moreover, we also find that the
collision and merging can play a vital role in tuning to a
librating-circulating mode of the giants. In Section 2, we reproduce
a scenario of migration and capture into a 2:1 resonance. In Section
3, we introduce the model and initial setup for our study. We
present the main results in Section 4. We conclude the outcomes in
Section 5.

\section{Migration and Captured into 2:1 Resonance}
As mentioned previously, the 2:1 resonant configurations are easily
achieved under convergent migration of two giant planets. Following
N-body damping method given by \citet{lee02}, we modified the
MERCURY package \citep{cha99} to examine the situation of migration
when the star mass is close to $1M_\odot$. As shown in Figure 1, the
two planets are placed on coplanar circular orbit similar to GJ 876
planetary system, with $a_2=2a_1=1\rmn{AU}$. The migration velocity
of the outer planet can be obtained by computing the coefficient
($\sim5\times10^{-5}~\rmn{yr}^{-1}$) of the right side of $(1)$ by
substituting $M_\star$ for $1M_{\odot}$. The eccentricity damping of
the outer planet is assumed to be
$\dot{e}_2/e_2=-100~|\dot{a}_2/a_2|$. The calculations are carried
out using a hybrid symplectic integrator in MERCURY package. When
the outer planet is forced to migrate toward the center star with
$\dot{a}_2/a_2=-5\times 10^{-5}~yr^{-1}$, the eccentricities of the
the two planets are excited rapidly, and then they are indeed
captured into 2:1 mean motion resonance with two resonant variables
$\theta_1$, $\theta_2$ librating about $0^\circ$, while the
eccentricities equilibrate gradually. The migration scenario is
plausible. The anti-symmetric, symmetric, and asymmetric
configurations of 2:1 mean motion resonances have been formed, which
has also been applied to model the formation of the GJ 876 and HD
82943 planetary systems \citep{lee02, lee04, lee06, Bea06} by
differential migration of planets with constant masses and initially
coplanar nearly-circular orbits. In addition, \citet{Bea06} further
pointed out that if they suffer an adiabatic process of a slow
migration for their orbits, the two giant can be finally captured
into in apsidal corotations. \citet{lee04} showed that some 2:1
resonant configurations may origin from an alternation of the
planetary mass ratio of two planets under migration or
multiple-planet scattering in packed systems. In the following, we
will show that both collision and scattering events among the
terrestrial bodies and the giant planets can yield the 2:1
librating-circulating configurations.

\begin{figure}
\includegraphics[width=8cm]{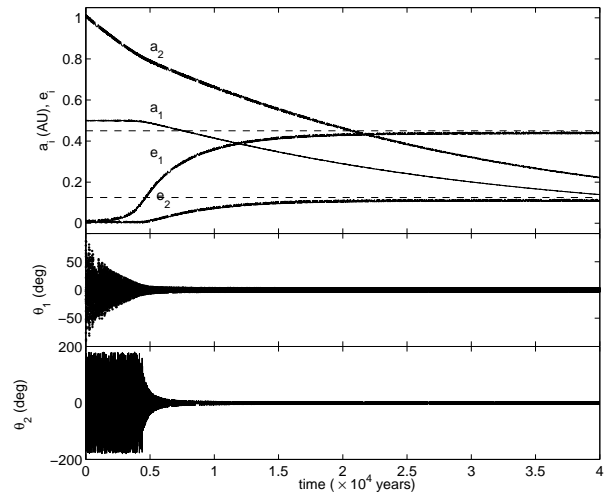}
\caption{Time evolution of the semi-major axes, eccentricities, and
resonant angles of two giant planets in the simulation. The initial
configuration for the giants was set on coplanar circular orbits
reference to GJ 876 system with $a_2=1\rmn{AU}$ and
$a_1=0.5\rmn{AU}$. The migration velocity
$\left|\dot{a_2}/a_2\right|$ is adopted to be $5\times
10^{-5}~\rmn{yr}^{-1}$, and the eccentricity damping of the outer
planet is assumed to be $\dot{e}_2/e_2=-100~|\dot{a}_2/a_2|$. The
eccentricities can be excited quickly, and then equilibrate
gradually after two giants are deeply locked into a symmetric 2:1
mean motion resonance, where both of the resonant variables
$\theta_1$, $\theta_2$ librate about $0^\circ$ with quite small
amplitudes.}
\end{figure}

\section{Model and  Numerical Setup}

In the late stage of planet formation, even though the region of the
inner disk contain little material after the giants formed, several
terrestrial planets may still form through accreting planetesimals
or embryos in a wider region of their feeding zones \citep{cha01,
ray04, ray05, ray06, Zha09}.  Another alternative scenario is that
the small bodies may be captured into mean motion resonances with
the giant planets under migration \citep{Mal00}. The shepherd
mechanism of the giants can also drive a swarm of planetesimals
toward the terrestrial region as the sources for further formation.
In this sense, it is reasonable to assume the system is gas-free,
and the migration of the giants has been stopped when the
terrestrial planets formed. It is supposed that there are a pair of
giants with commensurable orbits and several terrestrial planets
inside the region of the inner giant in the system. The planetary
configuration is much closer to physical scenario and convenient to
study the diversity of resonance scenario, through dynamical
evolution of the terrestrial planets.

In this work, we choose HD 128311 system as the basic framework for
the model. \citet{vog05} mentioned that the dynamical stable
solutions of the HD 128311 are not unique. Here we modify the
eccentricities of the two giants as our initial eccentricities in
order to simulate a little earlier state of the system (see Table
2). The other parameters for the giant planets are the same as
\citet{san06}, which are listed in Table 1. The modified orbital
parameters are dynamically stable at least over $10^6$ yr under
mutual gravitational interaction. In the work of \citet{san06}, the
semi-major axis of an additional small planet is set to be
$0.5\rmn{AU}$, about half of the semi-major axis of the inner planet
($\sim1.09\rmn{AU}$). As Lee et al. pointed out that a small
additional planet is easy to be captured into mean motion resonance
\citep{lee02, lee04}, it indeed happened before scattered for the
scenario of  the inner planet \citep{san06}. In our model, we follow
this strategy and place $2-4$ terrestrial planets about
$0.5\rmn{AU}$ to examine the evolution\footnote{As aforementioned,
\cite{ray04} show that a few terrestrial planets with masses close
to the Earth's mass can be formed within 1 AU. However, due to the
limitation of current observational precision, the Earth-like
planets similar to those in solar system residing in the terrestrial
region have not been discovered yet. It is likely that all the super
Earths recently discovered may be formed farther from the central
stars and then migrated to their present positions. This may imply
the difficulty in making two or more of such planets in nearby
orbits remain stable during the evolution. The adopted hypothesis
should be examined by future high-precision observations, to unveil
multiple stable terrestrial planets in the Habitable zones.}. We
consider the condition that there is no terrestrial planet or
planetary embryo between the orbits of two giants in the beginning,
indicating that they are entirely cleared off. The semi-major axes
of terrestrial planets are distributed in the range
$0.3-0.75\rmn{AU}$. The maximum value of the semi-major axis is a
little larger than that of the 2:1 resonance regime of the inner
giant at $\sim0.7\rmn{AU}$. The terrestrial planets are not
well-separated in initial trajectories, therefore, they can approach
each other and further merge into a larger body soon after the
orbits evolve.

The eccentricities of the terrestrial planets are randomly adopted
in the range of 0.1 and 0.3, which is comparable to those of
planetary formation \citep{cha01, ray04, ray06, Zha09}. Certain
initial eccentricities mean that the planetary orbits may intersect
in the beginning of evolution. The planetary masses also play an
important role in the colliding or scattering event. Consequently,
the terrestrial planets occupy a total mass of $10-15M_\oplus$, and
these bodies are placed inside the inner giant. As regards to the
specific mass of an individual terrestrial planet, we treat the
issue by following the route of unequal-mass or equal-mass. In the
former case, the mass of a terrestrial body is varied randomly among
$3-6M_\oplus$, and the total mass is required to meet the above
condition.  For equal-mass, the planetary mass is $5M_\oplus$,
$4M_\oplus$ and $3M_\oplus$, respectively, corresponding to the
number of 2, 3 and 4 for the terrestrial planets in the system. In
this instance, the overall mass is $10M_\oplus$ or $12M_\oplus$.
Moreover, the density of the giants and terrestrial planet is,
respectively, considered with respect to that of Jupiter and Earth
in solar system. Hence, we can obtain the size of each body, and the
collision and merging can be also taken into account in the run. In
each simulation, the terrestrial planets are in coplanar orbits
reference to the giant planets. The other angles of the mean anomaly
$M$ and the longitudes of periapse $\varpi$ are randomly distributed
between $0^{\circ}$ and $360^{\circ}$ for each orbit.

\begin{table}
\centering
\begin{minipage}{100mm}
\caption{Initial eccentricities of two giants.}
\begin{tabular}{@{}ccccc@{}}
\hline
&M1 Inner&M1 Outer&M2 Inner&M2 Outer \\
\hline
Eccentricity&0.15  & 0.46  & 0.05 & 0.01 \\
\hline
\end{tabular}
\end{minipage}
\end{table}

We prepare 120 simulations for two groups and use the hybrid
symplectic integrator \citep{cha99} in MERCURY package to integrate
all the orbits. The algorithm is capable of handling close
encounters among the bodies. In the calculations, we consider that
close encounters will take place while the bodies are separated by
not more than their 3 Hill radii. And the collision and merging
occurs when the minimum distance between any of the two objects is
equal to or less than the summation of their physical radii. In
addition, we adopt 4 days as  length of time step, which is about
one twentieth period of the terrestrial body residing at 0.3 AU.
Most of the simulations are carried out for $10^3-10^4$ yr.
Nevertheless, a few simulations are performed to extend on longer
integral time.

\section{Results}

The previous works \citep{san06, san07} on librating-circulating
configurations of the 2:1 mean motion resonance in the HD 128311 and
HD 73526 systems have reported that such resonance configuration can
be produced by scattering of an additional small planet through
orbital migration of a sudden stop. Our results also confirm their
conclusion. Moreover, we further show that collision and
merging\footnote{In the following, we refer to collision as a
merging event between a small terrestrial planet and a giant planet,
while merging specifically, means the circumstance between two small
terrestrial planets. MERCURY package models two planets
inelastically to merge together and produce a single new body by
conserving mass and total momentum.} can play a major role in
dynamical evolution and both mechanisms can modify the resonance
angles of the giants. The configuration engaged in the mean motion
resonances is varied or destructed. Concretely, in our results,
there are 16 simulations involved in inducing librating-circulating
resonance configuration, among which about $56\%$ are shaped by
mixed events of planet-planet scattering and merging (or collision).
In this section, we present  major results of several simulations
that illustrate the dynamical interaction of planet-planet
scattering and collisions. In Table 3 are listed the initial
conditions of terrestrial planets in four runs. Next we will address
the possibility to form librating-circulating configuration with
certain initials for terrestrial planets during dynamical evolution.

\begin{table}
\centering
\begin{minipage}{85mm}
\caption{Initial conditions of the terrestrial planets for 4 runs.}
\begin{tabular}{@{}cccccc@{}}
\hline Terrestrial\\planets\footnote{We use four digits following
the letter M to label each simulation. The first digit denotes the
models using different initial eccentricities of the giant planets,
as shown in Table 2. The next digit indicates an alternative
strategy as treating the individual masses of the terrestrial
planets in the simulation (the label 1, 2 respectively for unequal
and equal masses). The last two digits are used to stand for each
simulation. The letter T means terrestrial planets. In each run, we
set all terrestrial planets on coplanar orbits with respect to those
of the giant planets.} & $m(M_\oplus)$ & $a(\rmn{AU})$& $e$ &
$\varpi(deg)$ & $M(deg)$ \\
\hline
M1204~T1    &5.0&0.66&0.14&344&226 \\
~~~~~~~~~~T2&5.0&0.63&0.29&171&238 \\
M1205~T1    &5.0&0.54&0.15& 92&219 \\
~~~~~~~~~~T2&5.0&0.39&0.11&280&225 \\
M1227~T1    &3.0&0.48&0.12& 51&132 \\
~~~~~~~~~~T2&3.0&0.34&0.17& 63&203 \\
~~~~~~~~~~T3&3.0&0.54&0.27& 56&209 \\
~~~~~~~~~~T4&3.0&0.58&0.16&118&245 \\
M2120~T1    &3.7&0.39&0.12&  3& 39 \\
~~~~~~~~~~T2&5.9&0.57&0.10&144& 32 \\
~~~~~~~~~~T3&4.9&0.56&0.15&344&317 \\
\hline
\end{tabular}
\end{minipage}
\end{table}

\subsection{Planet-planet scattering}

The present observations show that the eccentricities of the
extrasolar planets can be much higher on average than those of the
solar system. Many potential mechanisms (such as secular
perturbations from a massive planetary companion or passing stars,
interactions of orbital migration via mean motion resonance, or
planet-planet scattering in the planetary systems, etc., see details
in \citealt{For08}) have been proposed to explain the origin of the
distribution of high eccentricities. Based on the realistic initial
conditions, \citet{For08} and \citet{Cha08} show that the
combination of planet-planet scattering and tidal circularization
may lead to form several close-in giants and reproduce the observed
eccentricity distribution. In addition, in the following we show
that planet-planet scattering can also shape the
librating-circulating configuration trapped in resonance.

\begin{figure*}
\includegraphics[width=8cm]{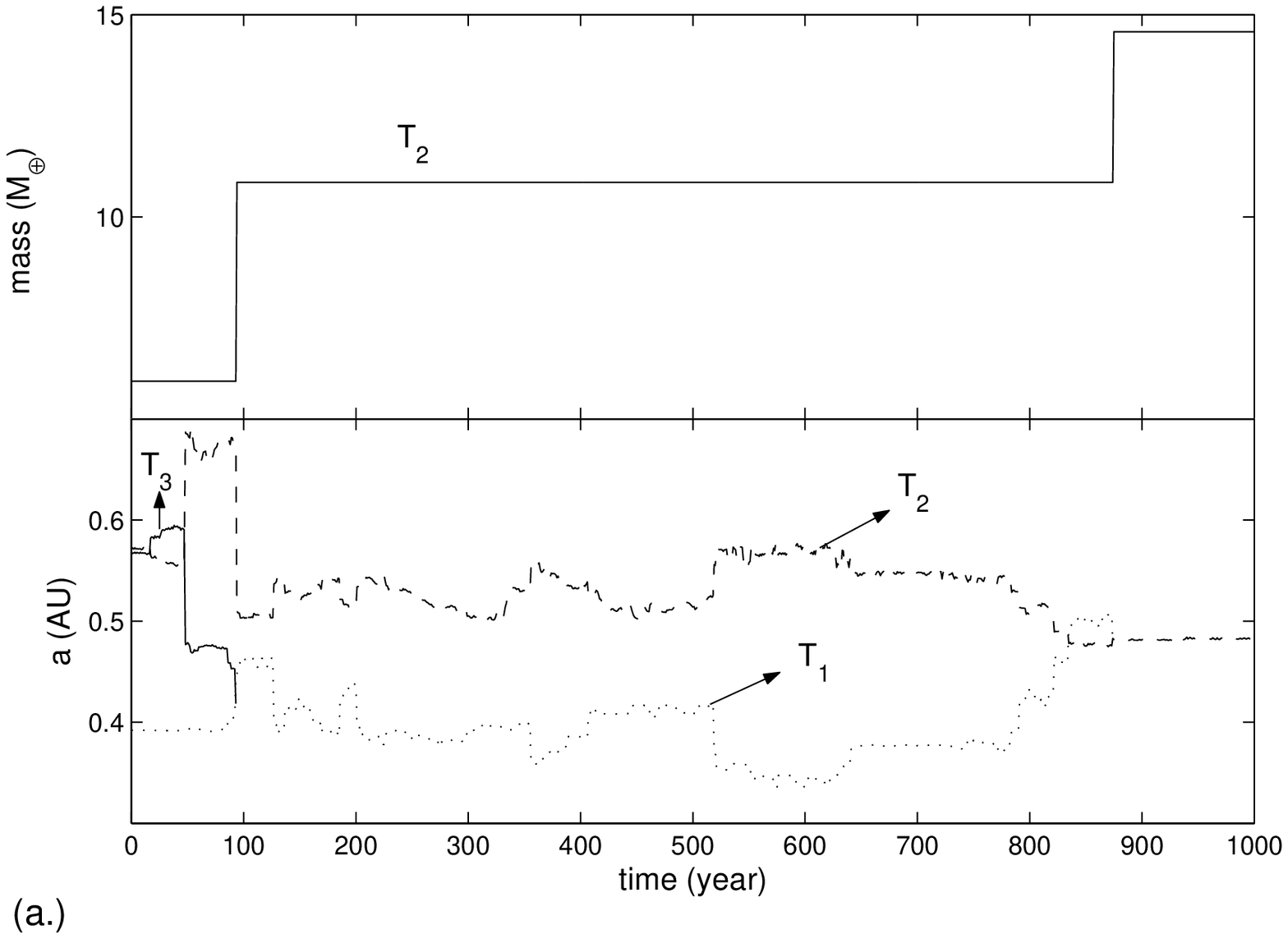}
\includegraphics[width=8cm]{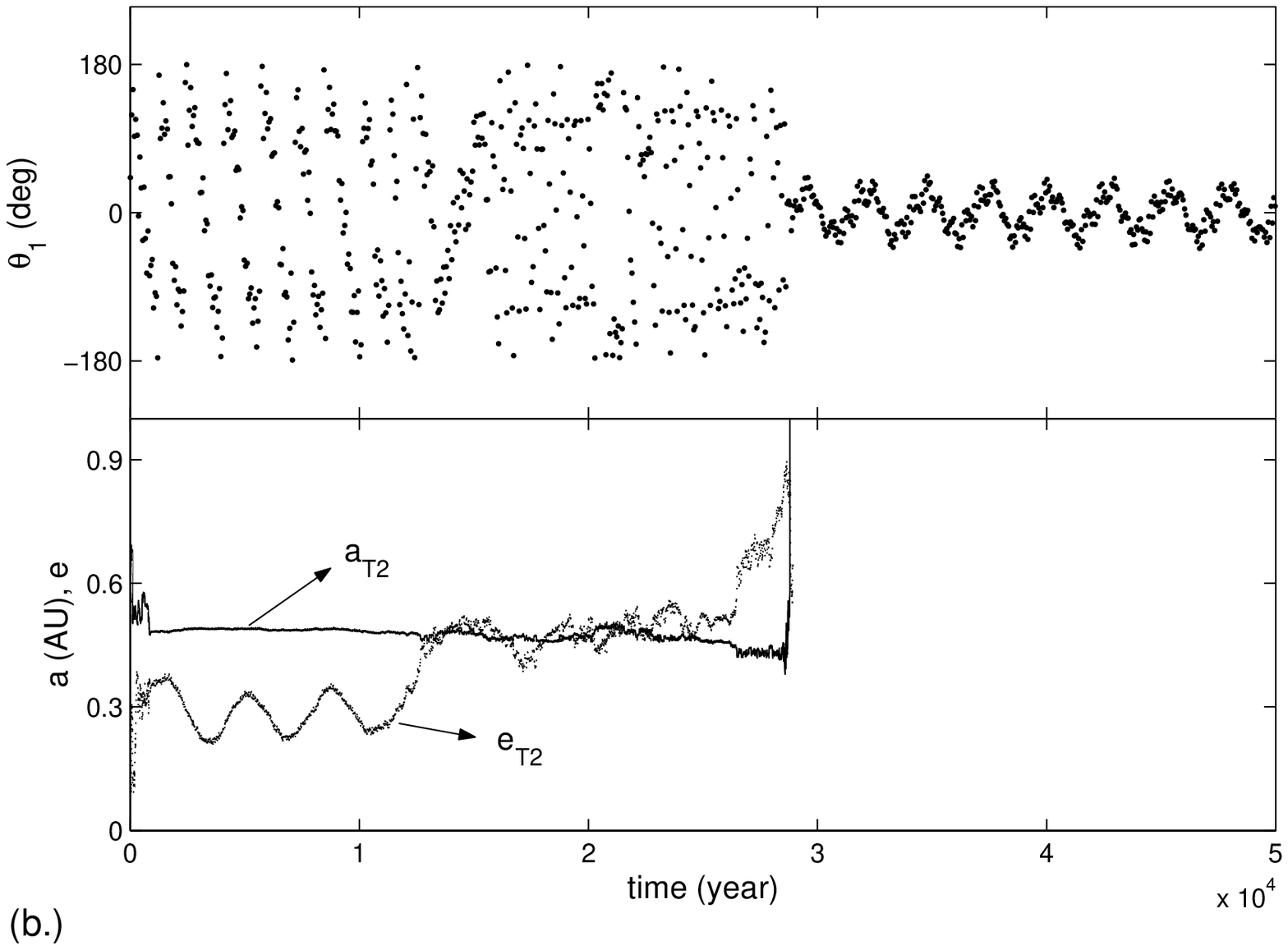}
\caption{(a) Time evolution of the mass of T2, and the semi-major
axes of T1, T2, T3 for simulation M2120. (b)Time evolution of the
resonance angle $\theta_1$,  semi-major axis and eccentricity of T2
in simulation M2120. T3 merges into T2 at about one hundred yr,
because of their very close initial orbits. After T1 merges into T2,
the eccentricity of 'Merger' T2 is greatly excited by gravitational
perturbation of the inner giant, and then T2 is scattered at $\sim$
$3\times10^4$ yr. The configuration is modulated to be a
librating-circulating stage after such scattering event.}
\end{figure*}

Librating-circulating configuration of mean motion resonance may be
formed by a single planet scattering event with moderate mass
\citep{san06}. Herein, we show an interesting example that is
consistent with such mechanism but related to a more complicated
physical scenario. Figure 2(a) illustrates the time evolution of the
mass of T2, and the semi-major axes of three additional terrestrial
planets in simulation M2120. From Table 3, we can clearly notice
that T2 and T3 are set close in the initial setup, and then they are
rapidly merged together into a larger body at about 100 yr. Due to
significant excitation of the eccentricities, the orbital crossings
of the terrestrial planets quickly occur and subsequently cause the
merging event between T2 and T3. In Figure 2(b), we can see the
'Merger' T2 remains in a stable orbit for about $10^4$ yr but its
eccentricity oscillates with larger amplitudes. And then the
consequence of stirring the eccentricity is that T2 is scattered at
$\sim$ $3\times10^4$ yr. Here comes the librating-circulating
configuration of the pair giants, where the resonance argument
$\theta_1$ modulates with moderate amplitudes about $0^\circ$, and
$\theta_2$ circulates as time evolves. The difference is that, in
our assumptions, the initial conditions of the two giants are simply
required to meet $a_2/a_1\simeq2/1$ (where the subscripts 1, 2
denote the inner and outer giant, respectively, hereinafter), so
their configuration after migration is not required to be exactly
trapped into 2:1 resonance, where two resonant angles are initially
both in circulating mode. In our model, we are not concerned about
how two giants planets move to the orbits close to the 2:1 resonance
through certain consequences of migration. But the initial setup we
require is that several terrestrial planets stay inside the region
of the inner giant in the gas-free disk after the termination of
migration for two giants.

We find that if the masses of the scattered terrestrial planets are
low, then the changes of the resonance configurations may reinstate
after some while. Generally, small planet scattering is capable of
increasing the amplitude of the resonance angles. It is very likely
that continuous scattering events resulting from small planets can
alter the configuration of the mean motion resonance. Figure 3
illustrates the results of the simulation of M1204. In this run,
there are two terrestrial planets with equal masses of $m_{T1}=
m_{T2}=5M_\oplus$ initially moving about the giants. From Table 3,
we know that, though their semi-major axes are originally
approximate to each other, the two terrestrial planets can escape
colliding from each other. Because the orbit of T1 is a bit broader
and less eccentric than that of T2, and the difference in the mean
longitudes of periapse is
$\Delta\lambda=\lambda_1-\lambda_2=161^\circ$ (where
$\lambda=\varpi+M$, $M$ is the mean anomaly), near  $180^\circ$. As
Figure 3 shows,  the 2:1 resonance of  two giants is initially in
symmetric configuration where
$(\theta_1,\theta_2)\approx(0^\circ,0^\circ)$. At the time the first
terrestrial planet T1 is scattered at 366 yr (marked by triangles),
the amplitudes of the resonance angles become enlarged, but the
apsidal corotation is maintained. Until the second terrestrial
planet is scattered at 593 yr, soon after the first scattering
event, the apsidal corotation is broken up. And the eccentricities
of two giants $e_1$ and $e_2$ fluctuate with large amplitudes, which
is similar to the HD 128311 and HD 73526 systems \citep{san06,
san07}. In the circumstances, our outcomes are well consistent with
those of \citet{san06}.

\begin{figure}
\includegraphics[width=8cm]{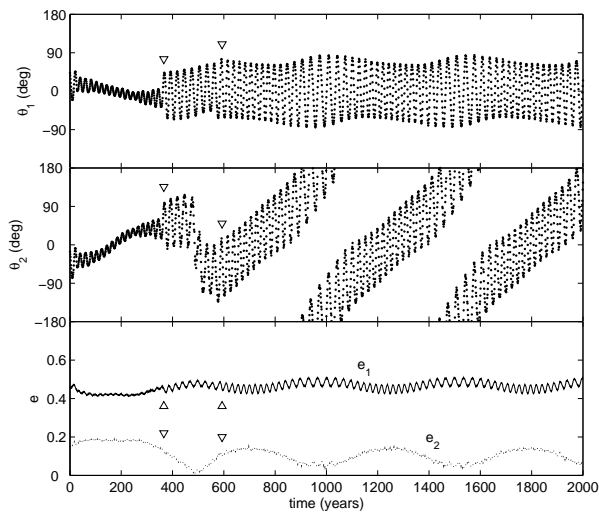}
\caption{Evolution of the 2:1 mean motion resonance variables
$\theta_1$, $\theta_2$, and the eccentricities of the two giants
$e_1$, $e_2$, for the simulation M1204. The sign triangles label the
time of two terrestrial objects that scattering happens. Obviously,
the resonance configuration after two sequent scattering events of
terrestrial planets turns into a librating-circulating state from an
original librating-librating mode of
$(\theta_1,\theta_2)\approx(0^\circ, 0^\circ)$.}
\end{figure}

\subsection{Merging and Collision}
As shown previously, the librating-circulating configuration can
arise from a planet-planet scattering of a moderate mass planet or
multiple continuous scattering for small terrestrial planets in the
various initial conditions. In this section, we will further focus
on the mixture role of merging and collision to generate a
librating-circulating mode in the dynamical evolution.

Figure 4 shows the results of M1227, which are involved in the very
complicated process of colliding and scattering events. In the
initial stage, there are 4 terrestrial planets in total, and each
terrestrial planet bears a pretty smaller mass of $3M_\oplus$. In
this scenario, it is more likely to observe multiple scattering or
colliding events (marked by upper letter '{\sl S}', '{\sl C}',
respectively in Fig.4) over timescale of the dynamical evolution.
From the figure, we can see that the inner giant is hit by T3, and
then T4 is scattered within 500 yr. However, we may notice that the
2:1 mean motion resonance configuration are not significantly
changed, where the first planet-planet scattering event or collision
with the inner giant before 500 yr do not modify the oscillating
status of two resonance arguments $\theta_1$ and $\theta_2$, due to
their smaller planetary masses ($3M_\oplus$). In the subsequent
evolution, T1 and T2 are merged together to form a larger
terrestrial body, and the eccentricity of the merger is gradually
excited to high values, which triggers the collision between a new
$6M_\oplus$ body and the outer giant. The second collision event of
the merged terrestrial planet and the outer giant occurs at about
3500 yr, which induce the two resonant angles $\theta_1$ and
$\theta_2$ to go from librating-librating phase into
librating-circulating phase. It is worthy to note that the
eccentricities of the giants undergo smaller oscillations. The
reason for the variations of the eccentricities is the conservation
of the total momentum, while the impactor loses the momentum and
transfers to the outer giant, which gains the momentum. Then the
eccentricity $e_2$ is slightly modulated if the semi-major axis
remains and further influences $e_1$ of the inner giant.

\begin{figure}
\includegraphics[width=8cm]{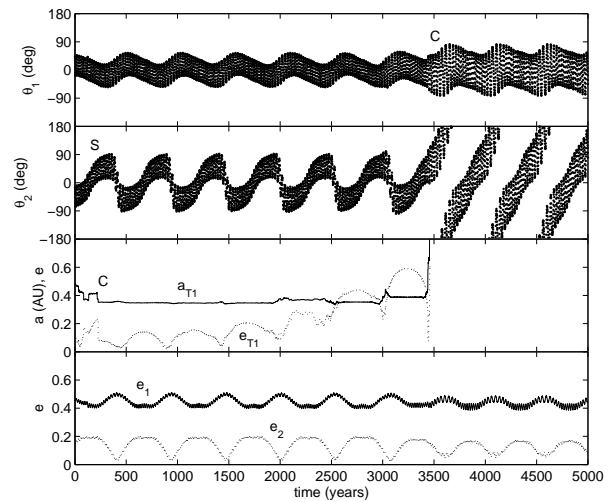}
\caption{Dynamical evolution of the 2:1 mean motion resonance
variables $\theta_1$ and $\theta_2$, and eccentricity $e_{T1}$
semi-major axis $a_{T1}$ of the terrestrial planet T1, and the
eccentricities $e_1$, $e_2$ of the two giants for the run of M1227.
The upper letters '{\sl C}' and '{\sl S}' denote a Collision or
Scattering event occurs at the corresponding time. Collision and
scattering events are observed in the simulation. However, the first
scattering event or collision with the inner giant before 500 yr
cannot change the vibration status of the resonance angles. The
configuration of mean motion resonance has been triggered to be a
librating-circulating mode at 3461 yr, because T1 and T2 are merged
into one new bigger terrestrial body, which collides with the outer
giant planet at about 3500 yr.}
\end{figure}

Figure 5 illustrates the evolution of simulation M1205. It is
similar to the case of M1227. However, the difference is that the
first '{\sl C}' denotes a merging event between T1 and T2 at about
500 yr. It is not difficult to understand that such merging scenario
for two terrestrial planets cannot alter the masses of giant
planets, and will not cause the variations of their semi-major axes
and eccentricities. However, subsequent evolution is to be expected
soon after the first collision. The second collision occurs between
T12 (as a merged larger body) and the inner giant at about 1300 yr.
In this case, two resonant arguments $\theta_1$ and $\theta_2$ vary
from a librating-librating phase into the librating-circulating
state. Similarly, both of the eccentricities for two giants are
slightly modulated. The above two figures show that a terrestrial
planet's colliding with either an inner or outer giant may finally
shape librating-circulating configuration of mean motion resonances
during the dynamical evolution.

\begin{figure}
\includegraphics[width=8cm]{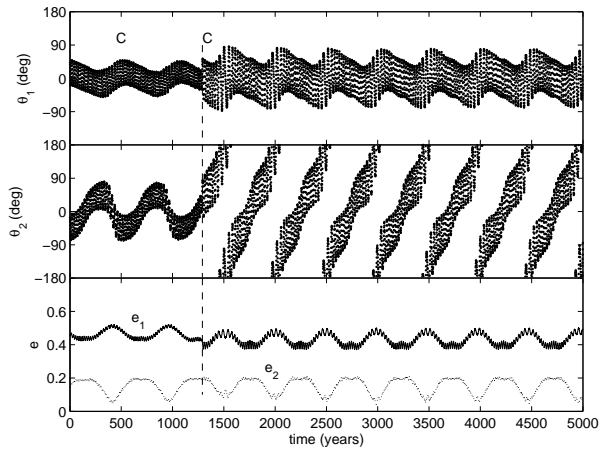}
\caption{Time Evolution of the simulation M1205. The first upper
letter '{\sl C}' denotes a merging event between T1 and T2. And the
second collision occurs between T12 (as a merged body) and the inner
giant. It is clear that colliding with inner giant can also form a
librating-circulating configuration engaged in  mean motion
resonances.}
\end{figure}

In addition, there are lots of simulations that show the destruction
of the systems, where one of the two giants is scattered faraway to
interstellar space or even entirely out of a planetary system. In
most of our runs, we notice that both scattering and colliding
events may coexist in the evolution. The planet-planet scattering
model \citep{For08,Cha08} can well explain the distribution of
relatively higher eccentricities of the extrasolar planets in
comparison with the observations, and further indicate that much
more planets can be excited in the eccentricity in the dynamical
evolution. In the late stage of planet formation, the evolution of
orbits is quite chaotic, because at that time the terrestrial
planets could frequently intersect with each orbit. In this sense,
it is very likely for terrestrial planets or massive planetary
embryos to suffer from gravitational scattering and collisions in
the system. In final, such hybrid mechanism may participate in
determining and forming a 2:1 librating-circulating resonant
configuration in the planetary system.

In order to examine the limiting conditions for the above mechanism
to work, we further performed over 100 sets of simulations to find
the minimum masses of the terrestrial planets, which convert
librating-circulating status for the giant planets. As for
planet-planet scattering case, in simulations related to M1204, we
find that when $m_{T1}= m_{T2}=3.8 M_\oplus$, the first terrestrial
planet T1 hit the inner giant at about 128 yr, and the second
terrestrial planet is scattered at 269 yr soon after the first
collision event, and the librating-circulating resonance
configuration is induced. In other simulations of this group, the
terrestrial planet masses $m_{T1}= m_{T2} < 3.8 M_\oplus$ cannot
trigger the librating-circulating resonance configuration. As for
merging and collision scenario, the additional runs of M1205 shows
that when $m_{T1}= m_{T2}= 2.5 M_\oplus$, a merging event between T1
and T2 occurred at about 368 yr, then the second collision occurs
between T12 (as a merged larger body) and the inner giant at about
671 yr, finally resulting in the librating-circulating configuration
of mean motion resonance. Also the simulations related to M1205 show
that the librating-circulating resonance configuration of two giant
planets cannot be formed, when the terrestrial planet masses
$m_{T1}= m_{T2}< 2.5 M_\oplus$. It is not difficult to understand
that the presented minimum masses of terrestrial planets have
different values in these simulations,  to alter the status of the
mean motion resonance configuration of the giants. This is because
such conversion of the resonant planetary configuration may mainly
not only depend on the terrestrial planet masses, but also rely on
their other initial conditions as well.

\section{Conclusions and Discussions}

In this work, we have investigated potential mechanisms to shape 2:1
librating-circulating resonance configuration by considering a
planetary system of two giants accompanying with few terrestrial
planets with coplanar orbits in the late stage of planet formation.
In the model, the system is considered to be gas-free, and two
giants with commensurable orbits stop migrating and several
terrestrial planets have formed inside the inner giant. The
configuration is much closer to real planetary formation scenario
according to numerical simulations \citep{cha01, ray04, Zha09}. In
conclusion, we summarize the main results as follows.

In the late stage of planetary formation, planet-planet scattering
or colliding among planetesimals and embryos can frequently occur.
Our results show that not only a single planetary scattering of a
terrestrial planet with a moderate mass can result in the
librating-circulating configuration, but several continuous
planetary scattering with rather smaller terrestrial masses is also
at work. Additionally, if  two giant planets are initially engaged
in a 2:1 symmetric resonance and their eccentricities oscillate with
large amplitudes, the collisions arising from the giants and other
small bodies may change librating amplitudes of the resonance angles
during the evolution. If the apsidal corotation is disintegrated,
the configuration may turn into a librating-circulating status.
Obviously, the more mass of a perturbing terrestrial body may have
much greater influence on the commensurable giant planets. In most
simulations, colliding and scattering events can be found and they
can increase or decrease the fluctuation in the amplitude of the
resonant angles, even dramatically destroy the whole system. In a
word, the librating-circulating configuration of mean motion
resonance is likely to generate by a mixed mechanism of colliding
and scattering.

In addition, the librating-circulating configuration could be
generated through long-term evolution of planetary formation. A
librating-circulating configuration trapped in 3:2 mean motion
resonance \citep{Zha09} is unveiled during the formation of the
terrestrial planets, which is similar to the newly-discovered
planetary system HD 45346 \citep{cor09}. They show that the two
terrestrial planets were formed within 50 Myr (see their Figure 6
for details), and after that time the frequent orbital crossings of
them and their interaction with the inner giant planet may finally
lead to a capture of a 3:2 resonance. Such resonant configuration is
believed to hold over hundreds of Myrs.  However, more resonant
configurations for less massive planets are expected to reveal by
future higher accuracy space-based projects in search for
terrestrial planets (such as TPF, Darwin, SIM). The innovative
findings will encourage one to more carefully study their dynamics
and origin.

\section*{Acknowledgments}
We thank the anonymous referee for useful comments and suggestions
that helped to improve the contents. We are grateful to M. Mayor, D.
Bennett, M. H. Lee and W. Kley for insightful discussions. This work
is financially supported by the National Natural Science Foundation
of China (Grants 10973044, 10833001, 10573040, 10673006, 10233020),
the joint project by the Academy of Finland and NSFC (Grant
10911130220), the Natural Science Foundation of Jiangsu Province,
and the Foundation of Minor Planets of Purple Mountain Observatory.


\begin{thebibliography}{}

\bibitem[\protect\citeauthoryear{Beaug{\'e} et al.}{2003}]{Bea03}
Beaug{\'e} C., Ferraz-Mello S., Michtchenko T. A., 2003, ApJ, 593,
1124


\bibitem[\protect\citeauthoryear{Beaug{\'e} et al.}{2006}]{Bea06}
Beaug{\'e} C., Michtchenko T. A., Ferraz-Mello S., 2006, MNRAS, 365,
1160

\bibitem[\protect\citeauthoryear{Calvet et al.}{2005}]{cal05}
Calvet N., D'Alessio P., Watson D. M., et al., 2005, ApJ, 630, L185

\bibitem[\protect\citeauthoryear{Chambers}{1999}]{cha99}
Chambers J. E., 1999, MNRAS, 304, 793

\bibitem[\protect\citeauthoryear{Chambers}{2001}]{cha01}
Chambers J. E., 2001, Icarus, 152, 205


\bibitem[\protect\citeauthoryear{Chatterjee et al.}{2008}]{Cha08}
Chatterjee S., Ford E. B., Matsumura S., Rasio F. A., 2008, ApJ,
686, 580


\bibitem[\protect\citeauthoryear{Correia et al.}{2009}]{cor09}
Correia A. C. M., Udry S., Mayor M., et al., 2009, A\&A, 496, 521

\bibitem[\protect\citeauthoryear{D'Alessio et al.}{2005}]{dal05}
D'Alessio P., Hartmann L., Calvet N., et al.,  2005, ApJ, 621, 461

\bibitem[\protect\citeauthoryear{Ford \& Rasio}{2008}]{For08}
Ford E. B., \& Rasio F. A., 2008, ApJ, 686, 621

\bibitem[\protect\citeauthoryear{Gayon \& Bois}{2008}]{Gay08}
Gayon J., Bois E., 2008, A\&A, 482, 665

\bibitem[\protect\citeauthoryear{Go{\'z}dziewski \& Maciejewski}{2001}]{Goz01}
Go{\'z}dziewski K., \& Maciejewski A.~J.\ 2001, ApJL, 563, L81

\bibitem[\protect\citeauthoryear{Hadjidemetriou}{2002}]{Haj02}
Hadjidemetriou J.~D., 2002, CeMDA, 83, 141

\bibitem[\protect\citeauthoryear{Hadjidemetriou \& Psychoyos}{2003}]{Haj03}
Hadjidemetriou J. D., \& Psychoyos D. 2003, in Galaxies and Chaos,
ed. G. Contopoulos \& N. Voglis (Berlin: Springer), 412

\bibitem[\protect\citeauthoryear{Hadjidemetriou}{2006}]{Haj06}
Hadjidemetriou J.~D., 2006, CeMDA, 95, 225

\bibitem[\protect\citeauthoryear{Ji et al.}{2003a}]{Ji03a}
Ji J., Kinoshita H., Liu L., Li G., Nakai H., 2003, CeMDA, 87, 113

\bibitem[\protect\citeauthoryear{Ji et al.}{2003b}]{Ji03b}
Ji J., Liu L., Kinoshita H., Zhou J., Nakai H., Li G., 2003, ApJ,
591, L57

\bibitem[\protect\citeauthoryear{Ji et al.}{2003c}]{Ji03c}
Ji J., Kinoshita H.,  Liu , L., Li, G.,  2003, ApJ, 585, L139

\bibitem[\protect\citeauthoryear{Kley, Peitz,\& Bryden}{2004}]{Kle04}
Kley W., Peitz J., Bryden G., 2004, A\&A, 414, 735

\bibitem[\protect\citeauthoryear{Kley et al.}{2005}]{Kle05}
Kley W., Lee M.~H., Murray N.,  Peale S.~J., 2005, A\&A, 437, 727


\bibitem[\protect\citeauthoryear{Lee \& Peale}{2002}]{lee02}
Lee M. H., Peale S. J., 2002, ApJ, 567, 596

\bibitem[\protect\citeauthoryear{Lee}{2004}]{lee04}
Lee M. H., 2004, ApJ, 611, 517

\bibitem[\protect\citeauthoryear{Lee et al.}{2006}]{lee06}
Lee M. H., Butler R. P., Fischer D. A., Marcy G. W., Vogt S. S.,
2006, ApJ, 641, 1178

\bibitem[\protect\citeauthoryear{Lee \& Thommes}{2009}]{lee09}
Lee, M.~H., \& Thommes, E.~W., 2009, ApJ, 702, 1662


\bibitem[\protect\citeauthoryear{Malhotra et al.}{2000}]{Mal00}
Malhotra R., Duncan M. J.,  Levison H., 2000, in Protostars and
Planets IV, ed. V. Mannings, A. P. Boss, S. S. Russell(Tucson: Univ.
Arizona Press), 1231


\bibitem[\protect\citeauthoryear{Marcy et al.}{2001}]{mar01}
Marcy G. W., Butler R. P., Fischer D. A., Vogt S. S., Lissauer J.
J., Rivera  E. J., 2001, ApJ, 556, 296

\bibitem[\protect\citeauthoryear{Masset et al.}{2006}]{mas06}
Masset F. S., Morbidelli A., Crida A., Ferreira J., 2006, ApJ, 642,
478


\bibitem[\protect\citeauthoryear{Mayor et al.}{2004}]{may04}
Mayor M., Udry S., Naef D., Pepe F., Queloz D., Santos D. C., Burnet
M., 2004, A\&A, 415, 391


\bibitem[\protect\citeauthoryear{Raymond, Quinn \& Lunine}{Raymond et al.}{2004}]{ray04}
Raymond  S. N., Quinn  T., Lunine  J. I., 2004, Icarus, 168, 1

\bibitem[\protect\citeauthoryear{Raymond, Quinn \& Lunine}{Raymond et al.}{2005}]{ray05}
Raymond  S. N., Quinn  T., Lunine  J. I., 2005, ApJ, 632, 670

\bibitem[\protect\citeauthoryear{Raymond, Quinn \& Lunine}{Raymond et al.}{2006}]{ray06}
Raymond  S. N., Quinn  T., Lunine  J. I., 2006, Icarus, 183, 265


\bibitem[\protect\citeauthoryear{S\'{a}ndor \& Kley}{2006}]{san06}
S\'{a}ndor  Z., Kley  W., 2006, A\&A, 451, L31

\bibitem[\protect\citeauthoryear{S\'{a}ndor, Kley \& Klagyivik}{S\'{a}ndor et al.}{2007}]{san07}
S\'{a}ndor  Z., Kley  W., Klagyivik  P., 2007, A\&A, 472, 981 28

\bibitem[\protect\citeauthoryear{Tinney et al.}{2006}]{tin06}
Tinney  C. G., Butler  R. P., Marcy G. W., et al., 2006, ApJ, 2006,
647, 594


\bibitem[\protect\citeauthoryear{Vogt et al.}{2005}]{vog05}
Vogt  S. S., Butler  R. P., Marcy, G. W., et al., 2005, ApJ, 632,
638

\bibitem[\protect\citeauthoryear{Vogt et al.}{2009}]{vog09}
Vogt S. S., et al., 2009, ApJ, submitted, [arXiv:0912.2599]

\bibitem[\protect\citeauthoryear{Voyatzis et al.}{2009}]{voy09}
Voyatzis  G., Kotoulas  T.,  Hadjidemetriou  J.~D.,  2009, MNRAS,
395, 2147

\bibitem[\protect\citeauthoryear{Ward}{1997}]{war97}
Ward  W. R., 1997, Icarus, 126, 261

\bibitem[\protect\citeauthoryear{Zhang \& Ji}{2009}]{Zha09}
Zhang  N.,  Ji  J.,  2009,  Science  in  China  Series G , 52(5),
794

\end{thebibliography}
\end{document}